\documentclass[twocolumn,showpacs,preprintnumbers,amsmath,amssymb,groupedaddress,prl]{revtex4}

\usepackage{graphicx}
\usepackage{dcolumn}
\usepackage{bm}
\usepackage{epsfig}

\def\e{\begin{equation}}
\def\f{\end{equation}}
\def\=#1{\overline{\overline #1}}

\def\-#1{{\bf #1}}
\def\.{\cdot}

\def\vec#1{{\bf #1}}

\begin{document}

\title{Sub-wavelength imaging at optical frequencies using canalization regime}

\author{Pavel A. Belov}
\affiliation{Queen Mary College, University of London, Mile End
Road, London, E1 4NS, United Kingdom}

\author{Yang Hao}
\affiliation{Queen Mary College, University of London, Mile End
Road, London, E1 4NS, United Kingdom}

\begin{abstract}
Imaging with sub-wavelength resolution using a lens formed by
periodic metal-dielectric layered structure is demonstrated. The
lens operates in canalization regime as a transmission device and it
does not involve negative refraction and amplification of evanescent
modes. The thickness of the lens have to be an integer number of
half-wavelengths and can be made as large as required for ceratin
applications, in contrast to the other sub-wavelength lenses formed
by metallic slabs which have to be much smaller than the wavelength.
Resolution of $\lambda/20$ at 600 nm wavelength is confirmed by
numerical simulation for a 300 nm thick structure formed by a
periodic stack of 10 nm layers of glass with $\epsilon=2$ and 5 nm
layers of metal-dielectric composite with $\epsilon=-1$. Resolution
of $\lambda/60$ is predicted for a structure with same thickness,
period and operating frequency, but formed by 7.76 nm layers of
silicon with $\varepsilon=15$ and 7.24 nm layers of silver with
$\varepsilon=-14$.
\end{abstract}


\maketitle

The possibility of imaging with sub-wavelength resolution was first
reported by Pendry in 2000 \cite{Pendrylens}. It was shown that the
slab of left-handed material \cite{Veselago}, a medium with both
negative permittivity and permeability, can create images with
nearly unlimited resolution. This idea impeached validity of
classical restriction on resolution of imaging systems: diffraction
limit and became a starting point for creation of new research area
of metamaterials \cite{Smithreview}, artificial media possessing
extraordinary electromagnetic properties usually not available in
the natural materials. The idea of Pendry's perfect lens is based on
such exotic phenomena observable in left-handed media as backward
waves, negative refraction and amplification of evanescent waves.
The far-field of a source is focused due to effects of backward
waves and negative refraction. The near field of the source, which
contains sub-wavelength details, is recovered in the image plane
because of the amplification of evanescent modes in the slab.
Currently, the samples of left-handed materials are created only in
microwave region \cite{ShelbyScience}. The creation of left-handed
materials at THz frequencies and in optical range meets with
problems related to the difficulty in getting required magnetic
properties \cite{YenTHZ,LindenTHZ} which have to be created
artificially. In the absence of magnetic properties the lenses
formed by materials with negative permittivity only (for example,
silver at optical frequencies) are still capable to create images
with sub-wavelength resolution, but the operation is restricted to
$p$-polarization only and the lens has to be thin as compared to the
wavelength \cite{Pendrylens}. This idea was confirmed by recent
experimental results \cite{Silversub} which demonstrated the reality
of sub-wavelength imaging using silver slabs in optical frequency
range. The resolution of such lenses is restricted by losses in the
silver, but this problem can be alleviated by cutting the slab into
the multiple thin layers \cite{Shamoninalayered,bundle1} and
introduction of active materials \cite{bundle2}. Unfortunately, at
the moment there is no recipe how to increase the thickness of such
lenses other than the introduction of artificial magnetism.

Competitive alternatives of left-handed media at optical frequencies
are photonic crystals \cite{PhotJMW,sakoda}. The negative refraction
effect in photonic crystals at the frequencies close to the band-gap
edges was reported by Notomi in \cite{NotomiPRB,NotomiOQE} and the
sub-wavelength imaging using flat lenses formed by photonic crystals
was demonstrated both theoretically
\cite{Allanglediag,Allanglelight,LuoNROE,ZhangANR,Zhangsec} and
experimentally \cite{Parimilens,Minexper}. Unfortunately, the
resolution of such lenses is strictly limited by period of the
crystal. This fundamental restriction was formulated and proven in
\cite{Subwavelength}. It means that it is impossible to get very
good sub-wavelength resolution using lenses formed by photonic
crystals since they operate in the regime when the wavelength in the
crystal is comparable with lattice period, but this wavelength can
not be shortened too much due to the lack of naturally available
high-contrast materials.

During studies of negative refraction and imaging in photonic
crystals it was noted that in the certain cases sub-wavelength
imaging happens due to the other principle than that in left-handed
materials. Actually, the negative refraction in photonic crystals is
observed either in forward wave regime, that usually corresponds to
the frequencies from the first propagation band
\cite{Allanglediag,Subwavelength,Zhangsec}, or in backward wave
regime, that corresponds to the second propagation band
\cite{Allanglelight,LuoNROE,ZhangANR}. The evidence of non-negative
refraction was reported by numerous authors
\cite{Chien,Li,Kuo,Chigrin,canal} for the crystals operating in the
first frequency band. The lenses formed by such crystals indeed
operate in the regime which does not involve negative refraction and
amplification of evanescent waves. This regime was called in
\cite{canal} as {\it canalization}. The slab of photonic crystal in
the canalization regime operates not like usual lens which focus
radiation into the focal point, it effectively works as a
transmission device which delivers sub-wavelength images from front
interface of the lens to the back one. The implementation of such a
regime becomes possible if the crystal has a flat iso-frequency
contour and the thickness of the slab fulfils Fabry-Perot condition
(an integer number of half-wavelengths) \cite{canal}. The flat
iso-frequency contour allows to transform all spatial harmonics
produced by the source, including evanescent modes, into propagating
eigenmodes of the crystal. This preserves sub-wavelength details of
the source which usually disappear with distance due to rapid
spatial decay of evanescent harmonics. These propagating eigenmodes
transmit image across the slab from the front interface to the back
one. The possible reflections from the interfaces are eliminated
with the help of Fabry-Perot resonance for transmission which in
this case holds for all incidence angles due to flatness of
iso-frequency contour.

The lenses operating in the canalization regime have same
restrictions on the resolution provided by periodicity as those
working in the left-handed regime: in order to get sub-wavelength
resolution it is required to have period of the structure to be much
smaller than the wavelength. In microwave region the canalization
regime with $\lambda/6$ resolution for $s$-polarization \cite{canal}
was implemented using an electromagnetic crystal formed by a lattice
of wires periodically loaded by capacitances \cite{lwPRE}. Such a
crystal has a resonant band-gap at very low frequencies (with
wavelength/period ratio $\lambda/a=14$) and dos not contain
high-contrast materials. The theoretical and numerical estimations
\cite{canal} were confirmed by experimental verification
\cite{pekkaexp} and $\lambda/10$ resolution was demonstrated. The
higher resolution can be achieved using loading by larger
capacitances, but the real implementations of these ideas meet with
such problems as strong losses and very narrow band-width of
operation.

An excellent possibility to realize the canalization regime for
$p$-polarization is provided by a wire medium, a material formed by
a lattice of parallel conducting wires
\cite{Rotmanps,Brown,pendryw,WMPRB}. This material supports very
special type of eigenmodes, so-called transmission line modes
\cite{WMPRB}, which transfer energy strictly along wires with the
speed of light and can have arbitrary transverse wave vector
components. It means that such modes correspond to completely flat
isofrequency contour which is the main requirement for implementing
the canalization regime. The detailed analytical, numerical and
experimental studies \cite{SWIWM} show that flat lenses formed by
the wire medium are capable to transmit sub-wavelength images with a
resolution equal to double period of the structure which can be made
as small as required. Effectively, such lenses work as
multi-conductor transmission lines (telegraph) or bundle of
sub-wavelength waveguides, which perform pixel-to-pixel imaging. It
is important that such lenses are matched to free space and do not
experience parasitic reflections from the interfaces. The
sub-wavelength imaging with $\lambda/15$ resolution at 1 GHz was
demonstrated in the work \cite{SWIWM}. The measured bandwidth of
operation is 18\%. Moreover, the structure is nearly not sensitive
to the losses. Thus, the lens can be made as thick as required. The
only restriction is that the thickness should be an integer number
of half-wavelengths (in order to fulfil Fabry-Perot condition).

The lens formed by wire medium is an unique sub-wavelength imaging
device for microwave frequencies where metals are ideally
conducting. At the higher frequencies including visible range such a
lens will not operate properly since the metals at these frequencies
have plasma-like behavior. In the present paper we propose a
different structure which can operate in the canalization regime at
optical frequency range. This is a sub-wavelength optical telegraph
which operates completely in the same principle as the slab of wire
medium at microwaves. It is known that the wire medium \cite{WMPRB}
can be described by spatially dispersive permittivity tensor of the
form \e \=\epsilon=\-x\-x+\-y\-y+\varepsilon \-z\-z,\ \varepsilon
(\omega,q_z)=1-\frac{k_0^2}{k^2-q_z^2},\f where $z$-axis is oriented
along wires, $k=\omega/c$ is wave number of the host medium,
$k_0=\omega_0/c$ is wave number corresponding to the plasma
frequency $\omega_0$ which depends on the lattice period and radius
of wires, $q_z$ is $z$-component of wave vector $\vec q$, $c$ is the
speed of light. For the transmission line mode $q_z=k$ and effective
permittivity $\epsilon$ becomes infinite. Thus, transmission line
modes effectively propagate in the medium with permittivity tensor
of the form \e \=\epsilon=\-x\-x+\-y\-y+\infty\:\-z\-z.
\label{eq:infe}\f

\begin{figure}[htb]
\centering \epsfig{file=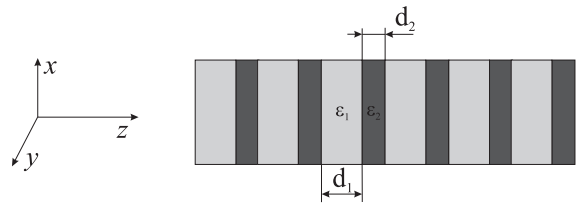, width=7.5cm} \caption{Geometry
of layered metal-dielectric metamaterial.} \label{geom1}
\end{figure}

In order to achieve in optical range the same properties as the wire
medium has at microwave frequencies it is required to find some
uniaxial optical material which has permittivity of the form
\ref{eq:infe}. Usually, it is assumed that in optical range it is
impossible to get very high values of permittivity. It is true for
natural materials, but for metamaterials, especially uniaxial, it is
not so. The high permittivity can be achieved in layered
metal-dielectric structures \cite{HIR}. Let us consider a layered
structure presented in Fig. \ref{geom1}.
Such a metamaterial can be described using permittivity tensor of
the form: \e
\=\epsilon=\varepsilon_\parallel(\-x\-x+\-y\-y)+\varepsilon_\perp
\-z\-z, \label{eq:epsl}\f where $$
\varepsilon_\parallel=\frac{\varepsilon_1d_1+\varepsilon_2d_2}{d_1+d_2},
\
\varepsilon_\perp=\left[\frac{\varepsilon_1^{-1}d_1+\varepsilon_2^{-1}d_2}{d_1+d_2}\right]^{-1}.$$
In order to get $\varepsilon_\parallel=1$ and
$\varepsilon_\perp=\infty$, and obtain a material with permittivity
tensor of the form \ref{eq:infe}, required for implementation of the
canalization regime, it is necessary to choose parameters of the
layered material so that $\varepsilon_1/\varepsilon_2=-d_1/d_2$ and
$\varepsilon_1+\varepsilon_2=1$. From the first equation it is clear
that one of the layers should have negative permittivity and thus,
the structure has to be formed by one dielectric layer and one
metallic layer. For example, one can choose $\varepsilon_1=2$,
$\varepsilon_2=-1$ and $d_1/d_2=2$, or $\varepsilon_1=15$,
$\varepsilon_2=-14$ and $d_1/d_2=15/14$.

Note, that no layered structure required for canalization regime can
be formed using equally thick layers $d_1=d_2$. The layered
metal-dielectric structures considered in
\cite{Shamoninalayered,bundle1,bundle2} have completely different
properties as compared to the structures considered in the present
paper. As it is noted in \cite{bundle2}, the structures with
$d_1=d_2$ and $\epsilon_1=-\epsilon_2$ (as in
\cite{Shamoninalayered,bundle1,bundle2}) correspond to
$\epsilon_\perp=0$ and $\epsilon_\parallel=\infty$, and operate as
an array of wires embedded into the medium with zero permittivity.
Such a structure can be considered as unmatched uniaxial analogue of
so-called material with zero-index of refraction \cite{Ziolzero}.
The absence of matching ($\mu=1$, but not $0$, as it is required)
causes strong reflections and restricts slab thickness to be thin.
In contrast to this case, in the canalization regime the reflections
from the slab are absent due to the Fabry-Perot condition which
holds for all angles of incidence.

\begin{figure}[htb]
\centering \epsfig{file=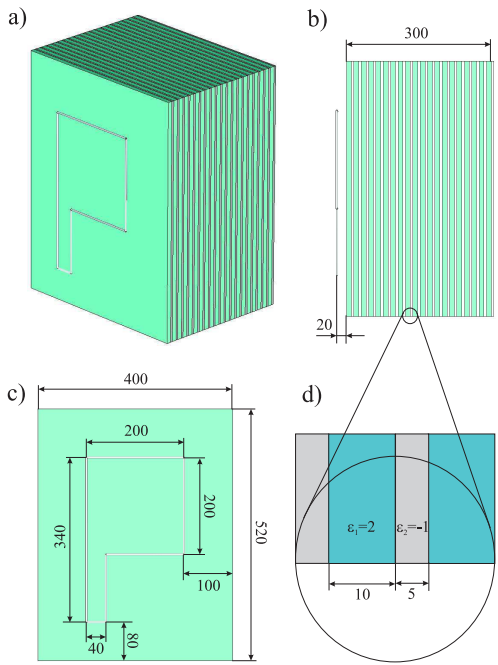, width=6.5cm} \caption{Geometry of
flat lens formed by a layered metal-dielectric metamaterial: a)
perspective view, b) side view, c) front view, d) small details. All
dimensions are in nm.} \label{geom2}
\end{figure}

\begin{figure}[tb]
\centering \epsfig{file=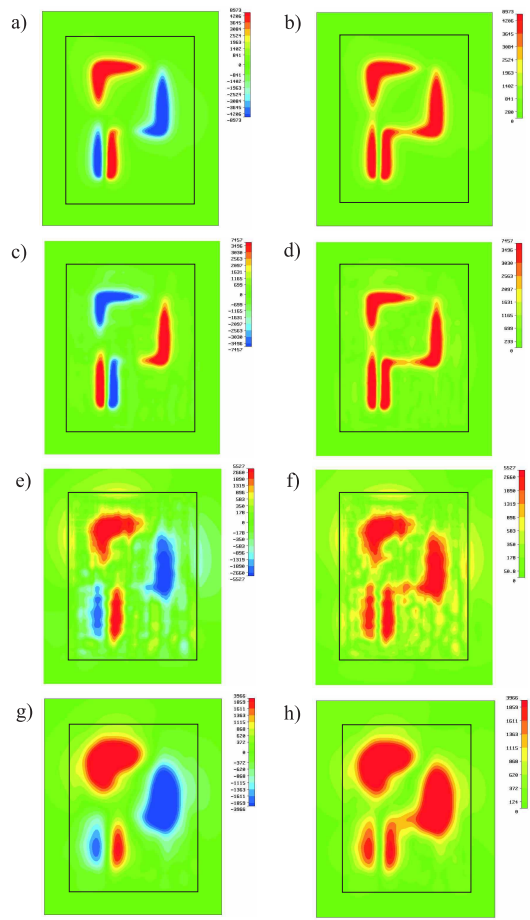, width=6.9cm} \caption{
Distributions of normal to the interface component of electrical
field: a) in free space, 20 nm from source, c) at the front
interface, e) at the back interface, g) in free space, 20 nm from
back interface; and their absolute values, b), d), f), h),
respectively.} \label{imag}
\end{figure}

In order to demonstrate how canalization regime can be implemented
using the suggested metal-dielectric layered structure, we present
results of numerical simulation using CST Microwave Studio package.
A sub-wavelength source (a loop of the current in the form of
P-letter) is placed at 20 nm distance from a 300 nm thick
multi-layer slab composed of 10 nm and 5 nm thick layers with
$\varepsilon_1=2$ and $\varepsilon_2=-1$, respectively. The detailed
geometry of the structure is presented in Fig. \ref{geom2}. The
wavelength of operation $\lambda$ is 600 nm. The field distributions
in the planes parallel to the interface of the lens plotted in Fig.
\ref{imag} clearly demonstrate imaging with 30 nm resolution
($\lambda/20$).
Figs. \ref{imag}.a,b show the field produced by the source in free
space at 20 nm distance. It is practically identical to the field
observed at the front interface of the lens, Figs. \ref{imag}.c,d.
It confirms that the reflections from the front interface are
negligibly small. Actually, the main contribution into reflected
field comes from diffraction at corners and wedges of the lens.
Figs. \ref{imag}.e,f show the field at the back interface of the
lens. The image is clearly visible, but it is a little bit distorted
by plasmon-polariton modes excited at the back interface. Being
diffracted into the free space this distribution forms an image
without distortions produced by plasmon-polariton modes, see Figs.
\ref{imag}.g,h presenting field distribution at 20 nm distance from
the back interface.

\begin{figure}[htb]
\centering \epsfig{file=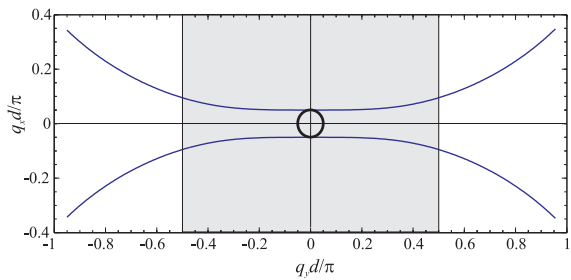, width=7.5cm}
\caption{Isofrequency contour of layered metal-dielectric structure
with $\varepsilon_1=2$, $\varepsilon_2=-1$, $d_1=10$ nm, $d_2=5$ nm,
$d=d_1+d_2=15$ nm for $\lambda=600$ nm. The region where the
dispersion curve is flat ($|q_xd/\pi|<0.5$) is shadowed.}
\label{disp}
\end{figure}

The resolution of the proposed layered lens is restricted by its
period $d=d_1+d_2$. The model of uniaxial dielectric \ref{eq:epsl}
is valid only for restricted range of wave-vector components. In
order to illustrate this we calculated isofrequency contour for the
layered structure under consideration treating it as 1D photonic
crystal and using an analytical dispersion equation available in
\cite{NefedovPRE}. The result is presented in Fig. \ref{disp}. While
$|q_xd/\pi|<0.5$ the isofrequency contour is flat, the homogenized
model \ref{eq:epsl} is valid and the lens works in the canalization
regime. The spatial harmonics which have $|q_xd/\pi|>0.5$ will be
lost by the lens and this defines $d/0.5\approx \lambda/20$
resolution. The calculation of isofrequency contour for the case of
$\varepsilon_1=15$, $\varepsilon_2=-14$, $d_1=7.76$ nm and
$d_2=7.24$ nm for the same wavelength of 600 nm revealed its
flatness for $|q_xd/\pi|>1.5$. This allows us to predict
$d/1.5\approx \lambda/60$ resolution for this case. Note, that the
similar restrictions on resolution by the periodicity are applicable
for the layered structures considered in \cite{bundle1,bundle2} in
spite of the authors' claims that resolution of such structures is
mainly limited by losses. The general study of limitations on
homogenization of periodic layered structures will be published
elsewhere.

In conclusion, we have demonstrated possibility of the imaging with
sub-wavelength resolution using the lens formed by a layered
metal-dielectric structure. The lens works in canalization regime as
a transmission device and does not involve negative refraction and
amplification of evanescent modes. The simulation was done for 300
nm thick structure comprising of 10 nm dielectric layers with
$\epsilon=2$ and 5 nm metal layers with $\epsilon=-1$ at wavelength
of 600 nm and resolution of $\lambda/20=30$ nm was shown. The metal
with $\epsilon=-1$ at 600 nm wavelength can be created by doping
some lossless dielectric by small concentration of silver which has
$\epsilon=-15$ at such frequencies, in the similar manner to the
ideas of works \cite{JavierPRL} or \cite{Shalaevmix}. Even more
promising resolution of $\lambda/60=10$ nm was predicted for the
layered structure comprising of 7.76 nm layers of dielectric with
$\varepsilon=15$ and 7.24 nm layers of metal with $\varepsilon=-14$.
The last structure can be constructed using silicon as dielectric
and silver as metal, but very accurate fabrication with error no
more than 0.05 nm will be required in order to get proper result.
The losses in silver in both cases are already reduced by operating
at rather long wavelength of 600 nm, but in accordance with our
estimations they are still high enough to destroy quality of the
sub-wavelength resolution. This problem can be solved by using of
active materials, for example doped silicon
\cite{bundle2,JavierPRL}.

\bibliography{silverlayer}
\end{document}